\documentclass[10pt,journal,compsoc]{IEEEtran}
\usepackage{color}

\ifCLASSOPTIONcompsoc
\usepackage[nocompress]{cite}
\else
\usepackage{cite}
\fi

\IEEEoverridecommandlockouts

\ifCLASSINFOpdf
\else
\fi

\def\ps@headings{%
	\def\@oddhead{\mbox{}\scriptsize\rightmark \hfil \thepage}%
	\def\@evenhead{\scriptsize\thepage \hfil \leftmark\mbox{}}%
	\def\@oddfoot{}%
	\def\@evenfoot{}}
\usepackage{epsfig, epsf, array, latexsym, graphics, multirow, color}
\usepackage{graphicx}
\usepackage{multirow}
\usepackage{booktabs,caption}
\usepackage[flushleft]{threeparttable}
\usepackage{amsmath}
\usepackage{kpfonts}
\usepackage{cuted}
\usepackage{flushend}
\usepackage[ruled]{algorithm}
\usepackage{algorithmicx}
\usepackage{algpseudocode}

\usepackage{helvet}         
\usepackage{courier}        
\usepackage{type1cm}        
\usepackage{makeidx}         
\usepackage{graphicx}        
\usepackage{multicol}        
\usepackage[bottom]{footmisc}

\newcommand {\mymarginpar}[1]{\marginpar{#1}}
\renewcommand {\marginpar}[1]{} 

\def\_{\rule{.3em}{.15ex}}      

\newcommand{\ls}[1]
{\dimen0=\fontdimen6\the\font
	\lineskip=#1\dimen0
	\advance\lineskip.5\fontdimen5\the\font
	\advance\lineskip-\dimen0
	\lineskiplimit=.9\lineskip
	\baselineskip=\lineskip
	\advance\baselineskip\dimen0
	\normallineskip\lineskip
	\normallineskiplimit\lineskiplimit
	\normalbaselineskip\baselineskip
	\ignorespaces
}

\newcommand {\bearn}{\begin{eqnarray*}}
	\newcommand {\eearn}{\end{eqnarray*}}
\newcommand {\barr}{\begin{array}}
	\newcommand {\earr}{\end{array}}

\def\defeq{\stackrel{\scriptstyle\rm def}{=}}

\newtheorem{definition}{Definition}
\newtheorem{property}[definition]{Property}
\newtheorem{proposition}[definition]{Proposition}
\newtheorem{lemma}[definition]{Lemma}
\newtheorem{theorem}[definition]{Theorem}
\newtheorem{corollary}[definition]{Corollary}
\newtheorem{example}[definition]{Example}
\newtheorem{remark}[definition]{Remark}
\newtheorem{conjecture}[definition]{Conjecture}
\newtheorem{assumption}[definition]{Assumption}

\newcommand {\benum} {\begin{enumerate}}
	\newcommand {\eenum} {\end{enumerate}}

\newcommand {\bdesc} {\begin{description}}
	\newcommand {\edesc} {\end{description}}

\newcommand {\bfig}[2] {\begin{figure}[htbp]
		\centerline {
			\epsfig{figure={#1},clip=,width={#2}}}}
	\newcommand {\brotatefig}[2] {\begin{figure}[htbp]
			\centerline {
				\epsfig{figure={#1},clip=,angle=-90,width={#2}}}}
			
\newcommand {\bfigfirst}[2] {\begin{figure}[h]
							\centerline {
								\setlength{\epsfxsize}{#2}
								\epsffile{#1}}}
\newcommand {\efig}[2]{ \caption{#2}
							\label{fig:#1}
						\end{figure}
						\mymarginpar{fig:#1}}
\newcommand {\erotatefig}[2]{ \caption{#2}
						\label{fig:#1}
					\end{figure}
					\mymarginpar{fig:#1}}
				\newcommand {\rfig}[1]{Figure \ref{fig:#1}}
				
				\newcommand {\btab}[1]{
					\begin{table}
						\centering
						\begin{tabular}{#1}}
						\newcommand {\etab}[3] {
						\end{tabular}
						\caption[#3]{#2}
						\label{tab:#1}
					\end{table}
					\mymarginpar{tab:#1}
					\vspace{.1in}}

				\newcommand {\btabular}[1]{\begin{center}
						\begin{tabular}{#1}}
						\newcommand {\etabular}{\end{tabular}
				\end{center}}
				
				\newcommand {\bdefin}[1]{\begin{definition}
						\mymarginpar{def:#1}
						\label{def:#1} }
					\newcommand {\edefin}       {\end{definition}}
				
				\newcommand {\bassum}[1]{\begin{assumption}
						\mymarginpar{ass:#1}
						\label{ass:#1} }
					\newcommand {\eassum}       {\end{assumption}}

				\newcommand {\bpro}[1]{\begin{property}
						\mymarginpar{pro:#1}
						\label{pro:#1} }
					\newcommand {\epro}   {\end{property}}

				\newcommand {\bprop}[1]{\begin{proposition}
						\mymarginpar{prop:#1}
						\label{prop:#1} }
					\newcommand {\eprop}       {\end{proposition}}

				\newcommand {\blem}[1]{\begin{lemma}
						\mymarginpar{lem:#1}
						\label{lem:#1} }
					\newcommand {\elem}   {\end{lemma}}

				\newcommand {\bthe}[1]{\begin{theorem}
						\mymarginpar{the:#1}
						\label{the:#1} }
					\newcommand {\ethe}   {\end{theorem}}

				\newcommand {\bcor}[1]{\begin{corollary}
						\mymarginpar{cor:#1}
						\label{cor:#1} }
					\newcommand {\ecor}   {\end{corollary}}

				\newcommand {\bax}[1]{\begin{axiom}
						\mymarginpar{ax:#1}
						\label{ax:#1} }
					\newcommand {\eax}       {\vspace{-.1in} \end{axiom}}

				\newcommand {\bconj}[1]{\begin{conjecture}
						\mymarginpar{conj:#1}
						\label{conj:#1} }
					\newcommand {\econj}       {\vspace{-.1in} \end{conjecture}}

				\newcommand {\bex}[2]{\vspace{.1in}
					\begin{example}
						\mymarginpar{ex:#1}
						{\bf #2}
						\label{ex:#1} \em}
					\newcommand {\eex}       {\end{example} \vspace{.3cm} }

				\newcommand {\brem}[1]{\begin{remark}
						\mymarginpar{rem:#1}
						\label{rem:#1} \em }
					\newcommand {\erem}   {\end{remark}}

				\newcommand {\beq}[1]{\mymarginpar{eq:#1}
					\begin{equation}
						\label{eq:#1} }
					
					\newcommand {\beqno}[1]{\mymarginpar{eq:#1}
						\begin{eqnarray}
							\nonumber}
						
						\newcommand {\eeq}       {\end{equation}}
					\newcommand {\eeqno}       { && \end{eqnarray}}
				\newcommand {\req}[1]{(\ref{eq:#1})}

				\newcommand {\bear}[1]{\mymarginpar{eq:#1}
					\begin{eqnarray}
						\label{eq:#1} }
					
					\newcommand {\bearno}[1]{\mymarginpar{eq:#1}
						\begin{eqnarray}
							\nonumber}
						
						\newcommand {\eear}{\end{eqnarray}}
					\newcommand {\eearno}{\end{eqnarray}}
				\newcommand {\bsel}{\left \{ \begin{array}{cl}}
					\newcommand {\esel}{\end{array} \right.}
				
				\newcommand {\bmat}[1]{\left [ \begin{array}{#1}}
					\newcommand {\emat}{\end{array} \right ]}

				\def\R{I\kern-0.30em R}
				\def\P{I\kern-0.30em P}
				
\newcommand {\bxfig}[2] {\begin{figure}[htbp]
						\centerline {
							\includegraphics[width=#2]{#1}}}
\newcommand {\brotatexfig}[2] {\begin{figure}[htbp]
							\centerline {
								\includegraphics[width=#2,angle=90]{#1}}}
\DeclareGraphicsExtensions{.pdf,.jpg,.png}

						\def\bff{{\mbox{\boldmath $f$}}}

						\def\bfx{{\mbox{\boldmath $x$}}}

						\def\bfsx{{\mbox{\boldmath\scriptsize $x$}}}

						\def\bfH{{\mbox{\boldmath $H$}}}
						
						\def\bfJ{{\mbox{\boldmath $J$}}}

\def\ie{{\em i.e.}\ }

\def\qs{q^S}
\def\qi{q^I}

\begin{document}

\title{A Degree Based Approximation of an SIR Model with
	Contact Tracing and Isolation}
\author{Duan-Shin Lee,~\IEEEmembership{Senior Member,~IEEE} Ting-Zhe Liu, 
	Ruhui Zhang, Cheng-Shang Chang~\IEEEmembership{Fellow,~IEEE}
	\IEEEcompsocitemizethanks{\IEEEcompsocthanksitem D.-S. Lee and C.-S. Chang are with the Institute of Communications
		Engineering, National Tsing Hua University, Hsinchu 30013, Taiwan, R.O.C.
		D.-S. Lee, R. Zhang and T.-Z. Liu are with the Institute of Computer Science, 
		National Tsing Hua University.
		(Email: lds@cs.nthu.edu.tw,  cschang@ee.nthu.edu.tw, huibrana@gapp.nthu.edu.tw,
		peter91015@gmail.com.)}
		\thanks{This research was supported in part by the Ministry of Science and Technology,
		Taiwan, R.O.C., under Contract 109-2221-E-007-093-MY2.}}
\IEEEtitleabstractindextext{
	\begin{abstract}
		In this paper we study a susceptible infectious recovered 
		(SIR) model with asymptomatic patients, contact tracing and isolation 
		on a configuration network. 
		Using degree based approximation, we derive a system of differential equations
		for this model.  This system can not be solved analytically.  We present
		an early-time analysis for the model.  The early-time analysis produces
		an epidemic threshold.  On one side of the threshold, the disease dies out quickly.
		On the other side, a significant fraction of population are infected.
		The threshold only depends on the parameters
		of the disease, the mean access degree of the network, and the fraction of asymptomatic 
		patients.  The threshold does not depend on the parameter of contact tracing and isolation
		policy.  We present an
		approximate analysis which greatly reduces computational complexity.   The nonlinear
		system derived from the approximation is not almost linear.  We present a 
		stability analysis for this system.  We 
		simulate the SIR model with contact tracing and isolation on five
		real-world networks.  Simulation results show that contact tracing
		and isolation are useful to contain epidemics.
\end{abstract}
}

\maketitle
	
\IEEEdisplaynontitleabstractindextext
\IEEEpeerreviewmaketitle

\IEEEraisesectionheading{\section{Introduction}\label{s:introduction}}	

\IEEEPARstart{I}{n} the early stage of an outbreak when there are not many infected individuals, contact tracing,
quarantine and isolation is an effective way to contain an infectious disease 
\cite{Hou2020, Aleta2020, Kucharski2020, Hellewell2020, Yan2009, Carlos2003, Gumel2004}. 
A succinct example is Taiwan.  Taiwan is in close proximity of China and has a large
portion of its populations residing and working in China.  It was expected that Taiwan suffered
from a major epidemic soon after the COVID-19 outbreak started in December 2019.  
It turns out that Taiwan had a relatively small number of infected individuals for
a long period of time until April 2022.  During this period of time, Taiwan 
successfully contained COVID-19 not by extreme measures such as city lock-downs, 
but by enforcing regulations on contact tracing, quarantine and isolation \cite{Wang2020}.

Contact tracing and isolation as a measure to contain COVID-19
has been studied by many researchers.   Hou et al. \cite{Hou2020} studied the
effectiveness of quarantining Wuhan city against COVID-19.
They used a mixed “susceptible exposed infectious recovered”
(SEIR) compartmental model, in which some infected patients are asymptomatic.
Hou et al. showed that, by reducing the contact rate of latent individuals, interventions such as
quarantine and isolation can effectively reduce the potential peak number of
COVID‐19 infections and delay the time of peak infection.
Aleta and et al. \cite{Aleta2020} built a synthetic population network
of the Boston metropolitan area in the United States from mobile devices and census data.
Aleta and et al. performed simulations to show that robust level of testing, 
contact-tracing and household
quarantine could keep the disease within the capacity of the healthcare
system while enabling the reopening of economic activities after a period of strict social distancing
control of the COVID-19 epidemic.  Kucharski et al. \cite{Kucharski2020} built a social
contact graph using BBC Pandemic data from 40162 UK participants.  The authors 
simulated the effect of a range of different testing, isolation, tracing, 
and physical distancing scenarios.
Hellewell \cite{Hellewell2020} established a contact graph and used simulations to quantify the
effectiveness of contact tracing and isolation of cases at controlling COVID-19.


To our knowledge all studies on contact tracing and isolation 
were based on empirical, statistical analysis or simulation studies.  We consider a
susceptible-infectious-recovered (SIR) model with contact tracing and isolation
on a social contact graph.  
In this paper we assume that the social contact graph is
a configuration model \cite{Newman2010} and apply a degree based approximation 
\cite{PS2001A, PS2001B, Barthelemy2004Velocity, Bar2005} to analyze this model.  
Suppose that the maximum degree of vertices in the graph is $K_{\mbox{max}}$.
The degree based approximation leads to a system of differential equations of size
$5(1+K_{\mbox{max}})$.  This system can not be solved analytically.  We present
an early-time analysis of the system assuming that time is small.  We obtain
an epidemic threshold.  On one side of the threshold, the disease dies out.
On the other side, a significant fraction of vertices are infected with the disease.
Interestingly, the threshold only depends on the infection rate and recovering rate
of the disease, the mean access degree of the network, and the fraction of asymptomatic 
patients.  The threshold does not depend on the parameter of contact tracing and isolation
policy.  This implies that contact tracing and isolation can not prevent a disease
from spreading widely.  However, it controls the size of the epidemic, if the disease
spreads widely. The early-time analysis is accurate only when the 
time is small.  We also propose
an approximation that reduces the complexity from $5(1+K_{\mbox{max}})$ equations to
five equations.  Through numerical studies, we show that the approximation method works
very well.  We perform a stability analysis for the nonlinear system derived from
the approximation.  This nonlinear system is not almost linear and the general stability
theory for almost linear systems can not be applied.  We present an analysis for the 
stability of this nonlinear system.  

From numerical studies, we show that the contact tracing
and isolation is effective in containing epidemics.  However, it comes at a cost.
With strict isolation policies, a significant fraction of susceptible population
is isolated.  This can be detrimental to the function of a society, as the 
work of isolated people (such as police, fire fighters, garbage collectors and etc.) must be 
taken over by someone else.  Our numerical results show that a strict 
policy can isolate a large fraction of susceptible individuals.
The configuration network that we model the social contact graph 
is mathematically convenient. However, it suffers
from a few disadvantages.  For instance, its clustering coefficients and degree correlations
are very small.  It is known that network clustering has a strong impact to the epidemic
\cite{House.2008,House.2009,Boguna2002,Boguna2003,Eguiluz2002}.
We simulate the SIR model with more realistic 
contact-tracing and isolation policies using five real-world networks.
We show that contact tracing and isolation can effectively contain the epidemic.

The outline of this paper is as follows.  In Section \ref{quarantine} we introduce
an SIR model with contact tracing and isolation.  We derive a set of differential
equations based on degree approximation.  In Section \ref{eta} we present an
early time analysis of the set of differential equations.  In Section \ref{aa}
we present an approximate analysis of the set of differential equations.
In Section \ref{nsr} we present the numerical and simulation results.
We present the conclusions of this paper in Section \ref{conclusions}.

\section{SIR model with contact tracing}\label{quarantine}

We consider a susceptible-infectious-recovered (SIR) model with contact
tracing and isolation.    For many diseases, including COVID-19, it is possible that an 
infected individual shows no symptoms and yet is
capable to transmit the disease to his/her contacts \cite{Rockx.2020, Du.2020,Kronbichler2020,Gao2021}.  
Such individuals are called asymptomatic individuals \cite{Mwalili2020}.  
An asymptomatic individual has
the ability to transmit the disease to his/her contacts.  However, an asymptomatic individual
shows no symptoms and can hardly be detected, unless he or she is traced because his or her
contacts show symptoms.  This feature makes the disease difficult to contain.

Suppose that a susceptible individual is infected with the disease.  
With probability $1-\alpha$, this individual is asymptomatic and is in the 
infected and asymptomatic state.  
He or she has the ability to infect his or her neighbors in the contact graph.  
On the other hand, with probability $\alpha$ this individual shows symptoms 
and is isolated right away. In addition, a fraction $\eta$ of his or her neighbors 
in the contact graph are quarantined and isolated.  An isolated individual has
a rate of $\gamma_1$ to be released from isolation.  A susceptible individual
has a rate of $\beta$ to contract the disease from an infected and asymptomatic neighbor.  
An infected individual, isolated or not, has a rate of $\gamma$ to enter a recovered state.
The following is a list of states, in which an individual can be.
\begin{itemize}
	\item {\em Susceptible}: An individual is susceptible if the individual is healthy and can
	contract the disease if the individual is in touch with an infected individual.
	\item {\em Infected and asymptomatic}: An individual is infected and shows no symptoms.
	\item {\em Susceptible and quarantined}: A susceptible individual is isolated, because
	this individual is in touch with an exposed individual.  An isolated and 
	susceptible individual can not contract the disease, since they are isolated.
	\item {\em Infected and quarantined}: An isolated individual is also infected.
	\item {\em Recovered}: An individual is in the recovered state, if the individual is 
	recovered from the disease, or is killed by the disease.
\end{itemize}
A schematic diagram that shows state transitions is displayed in \rfig{quarantine-sir}.

\bfig{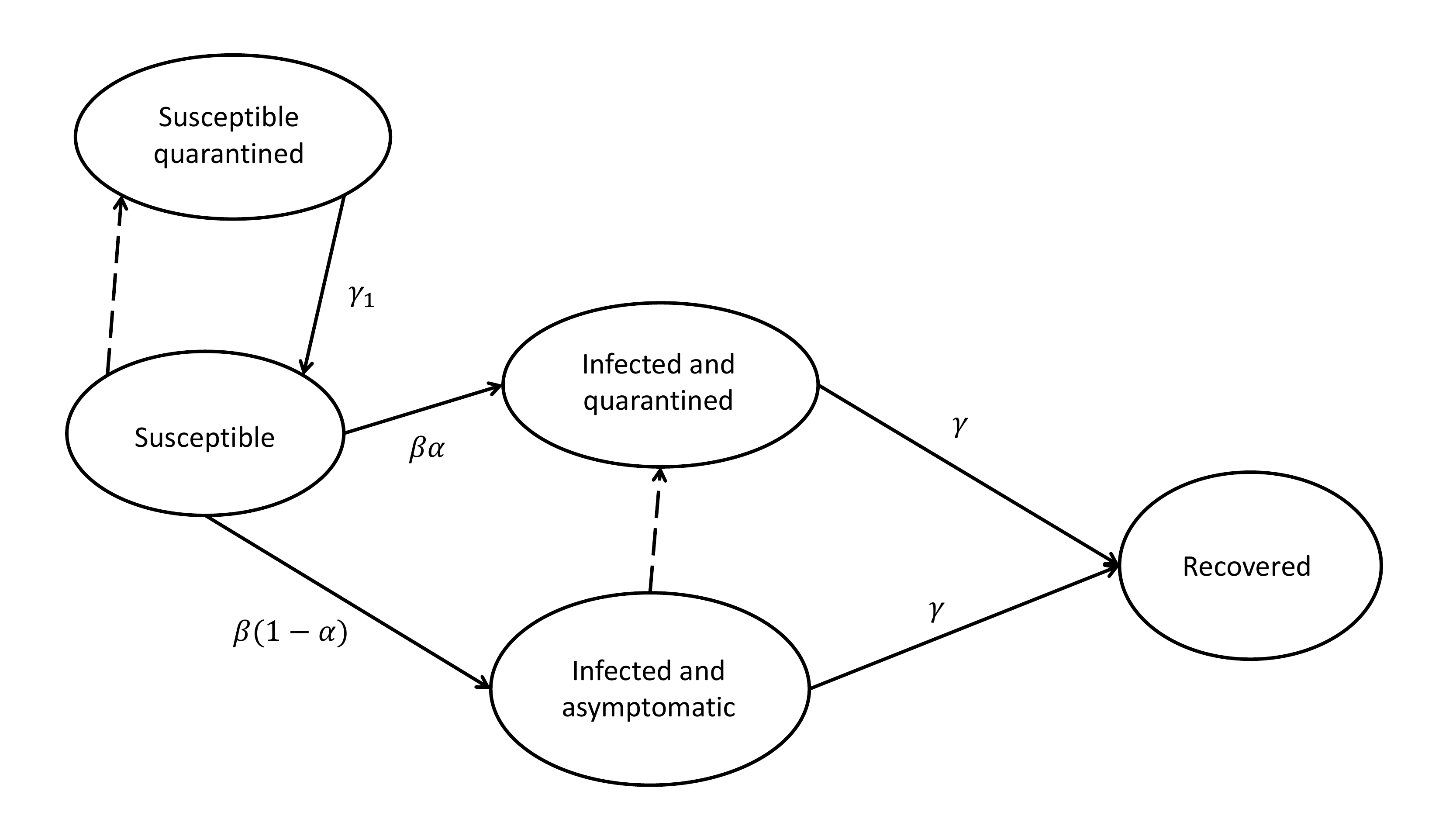}{3.5in}
\efig{quarantine-sir}{A transition diagram of an SIR model with contact tracing
	and isolation. Spontaneous transitions are shown in solid arrows.
	Dashed arrows denote transitions that are caused by spontaneous transitions.
	Specifically, as a susceptible individual becomes infected, a fraction $\eta$
	of his or her neighbors are tested.  His or her susceptible neighbors 
	(resp. asymptomatic neighbors) are isolated.  This causes
	a dashed transition from the susceptible state to the susceptible and quarantined
	state (resp. infected and quarantined state).  The transition rates of 
	spontaneous transitions are shown besides the corresponding arrows.}
\medskip

We apply a degree-based approximation to the SIR model with contact tracing and isolation.
Degree-based approximation is a well known technique to study epidemic spreading 
in a contact graph \cite{PS2001A, PS2001B, Barthelemy2004Velocity, Bar2005}.
The following is a list of variables that will be used in the analysis.
Note that excess degrees are defined to be the degree of a vertex reached by traversing
an edge.  The excess degree does not include the traversed edge.
\begin{itemize}
	\item $s_k(t)$: The fraction of nodes that are in state {\em Susceptible} at time $t$ among all
	nodes that have excess degree $k$.
	\item $\qs_k(t)$: The fraction of nodes that are in state {\em Susceptible and quarantined}
	at time $t$ among all nodes with excess degree $k$.
	\item $x_k(t)$: The fraction of nodes that are in state {\em Infected and asymptomatic}
	at time $t$ among all nodes with excess degree $k$.
	\item $\qi_k(t)$: The fraction of nodes that are in state {\em Infected and quarantined}
	at time $t$ among all nodes with excess degree $k$.
	\item $r_k(t)$: The fraction of nodes with excess degree $k$ that are in the 
	{\em recovered} state.
\end{itemize}
We remark that the quantities above satisfy an equality
\beq{equality}
s_k(t)+\qs_k(t)+x_k(t)+\qi_k(t)+r_k(t)=1
\eeq
for all $k$ and all $t$.

Let $\{p_k, k=0, 1, 2, \ldots\}$ be the degree distribution of a randomly selected vertex.
Let $\{w_k, k=0, 1, 2, \ldots\}$ be the excess degree distribution of a vertex reached
by traversing a randomly selected edge. 
That is, if one traverses along a randomly selected edge from one node to a neighbor
of the node, then $w_k$ is the probability that this neighbor has $k$ edges, 
not including the traversed edge.  
Let $g_0(x)$ and $g_1(x)$ be the probability generating functions of degree 
distribution $\{p_k, k\ge 0\}$ and excess degree distribution $\{w_k, k\ge 0\}$, 
respectively.  That is,
\begin{align}
	g_0(x) &= \sum_{k=0}^\infty p_k x^k, \label{def-g0}\\
	g_1(x) &= \sum_{k=0}^\infty w_k x^k.\label{def-g1}	
\end{align}
Let $K_0$ and $K_1$ denote the expectation of distributions of $\{p_k, k\ge 0\}$ and 
$\{w_k, k\ge 0\}$, respectively.  That is,
\begin{align}
	K_0 &=\sum_{k=0}^\infty k p_k = g_0'(1), \label{def-K0}\\
	K_1 &=\sum_{k=0}^\infty k w_k = g_1'(1). \label{def-K1}
\end{align}

Let
\beq{v}
v(t)=\sum_{k=0}^\infty w_k x_k(t).
\eeq
Then, $v(t)$ is the probability that a vertex reached by traversing a randomly 
selected edge is infected and not isolated.
We now derive a system of differential equations that link $s_k(t)$, $\qs_k(t)$,
$x_k(t)$, $\qi_k(t)$ and $r_k(t)$.
First,
\beq{s}
\frac{d}{dt}s_k(t) = -\beta k v s_k -\sum_{j=0}^\infty \left[\alpha\sum_{\ell=1}^\infty 
\beta \ell v s_\ell p_\ell \right] p_j j \eta s_k +\gamma_1 \qs_k.
\eeq
The left hand side of \req{s} is the rate of change in the fraction of susceptible nodes
with degree $k$.  The first term on the right side is the rate of change in the fraction of
susceptible nodes with degree $k$ due to infected but not isolated neighbors.  Each of
such a susceptible node has $k$ neighbors.  Each neighbor is infected
with probability $v(t)$ which is defined in \req{v}.  We thus have the first term 
on the right side.  The first term also appears in the degree based approximation
of the standard SI, SIR and SIS models.  We refer the reader to Newman \cite[p. 659, 665 and
671]{Newman2010}.  The fraction of vertices that have degree $k$, are isolated and susceptible is
$\qs_k(t)$. With rate $\gamma_1$, isolated and susceptible individuals are released from
isolation and return to susceptible state.  Thus, we have the third term on the right
side of \req{s}.  We now explain the second term.  The fraction of newly infected individuals
at time $t$ that have degree $k$ is $\beta k v s_k$.  Thus, the total fraction of newly infected
individuals at time $t$ is
\[
\sum_{k=1}^\infty \beta k v s_k p_k.
\]
Among those who are newly infected, a fraction $\alpha$ of individuals show symptoms and 
are quarantined and isolated.  A fraction $1-\alpha$ of individuals are asymptomatic
and are not isolated.  Thus, the fraction of newly infected individuals who show symptoms is
\[
\alpha\sum_{k=1}^\infty \beta k v s_k p_k.
\]
Those who show symptoms will be quarantined and isolated.
The fraction of newly infected individuals who are isolated and have degree $j$ is
\[
\left(\alpha\sum_{k=1}^\infty \beta k v s_k p_k\right)p_j.
\]
Each individual in the preceding quantity has $j$ neighbors.  
Each neighbor is isolated with probability $\eta$.  Thus, 
the fraction of newly isolated individuals is
\beq{total}
\sum_{j=0}^\infty\left(\alpha\sum_{k=1}^\infty \beta k v s_k p_k\right)p_j j \eta.
\eeq
These isolated neighbors are susceptible and have degree $\ell$
with probability $s_\ell(t)$.  Thus, the fraction of 
newly isolated and susceptible individuals who have degree $\ell$ is
\beq{total-s}
\sum_{j=0}^\infty\left(\alpha\sum_{k=1}^\infty \beta k v s_k p_k\right)p_j j \eta s_\ell(t).
\eeq
This is the second term on the right side of \req{s}.

With similar arguments one can derive differential equations for $\qs_k(t)$, $x_k(t)$,
$\qi_k(t)$ and $r_k(t)$. They are listed as follows.
\begin{align}
	\frac{d}{dt}\qs_k(t) &= \sum_{j=0}^\infty \left[\alpha\sum_{\ell=1}^\infty 
	\beta \ell v s_\ell p_\ell \right] p_j j \eta s_k -\gamma_1 \qs_k, \label{qs} \\
	\frac{d}{dt}x_k(t) &= (1-\alpha)\beta k v s_k-\sum_{j=0}^\infty \left[\alpha\sum_{\ell=1}^\infty 
	\beta \ell v s_\ell p_\ell \right] p_j j \eta x_k \nonumber\\
	&\quad-\gamma x_k, \label{x} \\
	\frac{d}{dt}\qi_k(t) &= \alpha\beta k v s_k+\sum_{j=0}^\infty \left[\alpha\sum_{\ell=1}^\infty 
	\beta \ell v s_\ell p_\ell \right] p_j j \eta x_k \nonumber\\
	&\quad-\gamma \qi_k, \label{qi} \\
	\frac{d}{dt}r_k(t) &= \gamma x_k+\gamma \qi_k.\label{r}
\end{align}
The derivation of the second term on the right side of (\ref{x}) and (\ref{qi}) is similar.
The fraction of newly isolated individuals is given in \req{total}.  The fraction of
isolated individuals who are infected (asymptomatically) and have degree $\ell$ is the 
quantity in \req{total} multiplied with $x_\ell(t)$.  This gives the second term on
the right side of (\ref{x}) and (\ref{qi}). We can simplify the second term on the
right side of \req{s} and (\ref{x}), respectively.  Eqs. \req{s} and (\ref{qs})-(\ref{qi})
become
\begin{align}
\frac{d}{dt}s_k(t) &= -\beta k v s_k -\alpha\beta v K_0 \eta\left[\sum_{\ell=1}^\infty 
\ell s_\ell p_\ell\right] s_k +\gamma_1 \qs_k, \label{s1}	 \\
\frac{d}{dt}\qs_k(t) &= \alpha\beta v K_0 \eta\left[\sum_{\ell=1}^\infty 
\ell s_\ell p_\ell\right] s_k -\gamma_1 \qs_k, \label{qs1}	 \\
\frac{d}{dt}x_k(t) &= (1-\alpha)\beta k v s_k-\alpha\beta v K_0 \eta\left[\sum_{\ell=1}^\infty 
\ell s_\ell p_\ell\right] x_k-\gamma x_k, \label{x1} \\
\frac{d}{dt}\qi_k(t) &= \alpha\beta k v s_k+\alpha\beta v K_0 \eta\left[\sum_{\ell=1}^\infty 
\ell s_\ell p_\ell\right] x_k-\gamma \qi_k, \label{qi1} 
\end{align}
where $K_0$ defined in (\ref{def-K0}) denotes the mean degree of a randomly selected vertex.
Unlike the differential equations for the classical SIR model which can be analytically
solved, the presence of the terms such as the second term on the right side of 
(\ref{s1}) makes (\ref{s1})-(\ref{qi1}) and (\ref{r}) very difficult.
They can only be solved numerically.  Once
$\{s_k(t), 0\le k< \infty\}$ are obtained, one can calculate
\beq{overall-r}
s(t)\defeq\sum_{k=0}^\infty p_k s_k(t).
\eeq
This is the probability that a randomly selected vertex is susceptible.
Probabilities that a randomly selected vertex is infected and quarantined, susceptible and quarantined,
infected but asymptomatic, or recovered can be obtained similarly.  

In Section \ref{eta}, we present an early-time analysis.
Numerical solution of the system is also difficult.  Suppose that we truncate
the excess degree distribution to $[0, K_{\mbox{\small max}}]$.  For each
degree value, there are five differential equations.
Thus, the dimension of the system in \req{s} and (\ref{qs})-(\ref{r}) is 
$5(1+K_{\mbox{\small max}})$.  We propose
an approximate analysis in Section \ref{aa} to reduce the numerical complexity.

\section{Early Time Analysis}\label{eta}

In this section we present an early time analysis of the system in \req{s}, (\ref{qs})-(\ref{r}).

We assume that initially there is small number of infected individuals and that most
individuals are susceptible.  That is, we assume that
\begin{align*}
&s_k(0) = 1-\epsilon,\quad x_k(0)  = \epsilon,\\	
&\qs_k(0) =\qi(0)=r(0)= 0, 
\end{align*}
for some small number $\epsilon \approx 0$.  Substituting $s_k(t)\approx 1-\epsilon$
in Eq. (\ref{x1}),  multiplying the two sides of (\ref{x1}) with $w_k$ and summing from
$k=0$ to infinity, we obtain
\[
\frac{dv}{dt}=[(1-\alpha)\beta g_1'(1) (1-\epsilon)-\gamma]v-
\alpha\beta \eta K_0^2(1-\epsilon) v^2.
\]
This differential equation is separable and can be rewritten as
\beq{separable}
\frac{dv}{v(c_2-c_1 v)}=dt,
\eeq
where 
\begin{align}
c_1 &= \alpha\beta \eta K_0^2 (1-\epsilon),	\label{c1}\\
c_2 &= (1-\alpha)\beta K_1 (1-\epsilon)-\gamma,\label{c2}
\end{align}
and $K_1=g_1'(1)$ is defined in (\ref{def-K1}).
The solution of \req{separable} is
\beq{v-sol}
v(t)=\frac{c_2 D_1 e^{c_2 t}}{1+c_1 D_1 e^{c_2 t}},
\eeq
where $D_1$ is a constant determined by the initial condition $v(0)=\epsilon$, \ie
\[
D_1=\frac{\epsilon}{c_2-c_1\epsilon}.
\]
If $c_2 < 0$, $v(t)\to 0$ as $t\to\infty$.  In this case, the epidemic dies down.
On the other hand, if $c_2 >0$, $v(t)\to c_2/c_1$ as $t\to\infty$.  In this case, a 
significant fraction of population will be infected with the disease.
Since $\epsilon$ is arbitrary, the condition for the epidemic to die down
is 
\[
(1-\alpha)\beta K_1 < \gamma
\]
and the condition for the epidemic to grow significantly is
\[
(1-\alpha)\beta K_1 > \gamma.
\]
Define 
\beq{R0}
R_0=\frac{(1-\alpha)\beta K_1}{\gamma}.
\eeq
The epidemic threshold is $R_0=1$, which separates the shrinking and growing regimes
of the epidemic.
$R_0$ is the basic reproduction number of the disease.  It is the expected number of
asymptomatic patients passed on by an infected individual during his/her infection.
It is interesting to see that $R_0$ depends only on the parameters of the disease, and not
on parameters $\eta$ and $\gamma_1$ of the contact tracing and isolation policy.

\section{Approximate Analysis}\label{aa}

In this section we present an approximation to reduce the dimension of the system
in \req{s} and (\ref{qs})-(\ref{r}).  We assume that 
\beq{s-approx}
s_k(t)=(u(t))^k
\eeq
for some unknown function $u(t)$.  Note that for the classical SI model and the
SIR model, \req{s-approx} holds for function $u(t)$ that depends on $v(t)$ and
$\{r_k(t), k\ge 0\}$, respectively.  We refer the reader to (17.60) and (17.87)
in Newman \cite{Newman2010}.  In our model, \req{s-approx} is only an
approximation.  However, it greatly simplifies the second term on the right side
of (\ref{s1}) that was caused by contact tracing and isolation.  

Define
\begin{align}
	\qs(t) &= \sum_{k=0}^\infty w_k \qs_k(t), \label{qsw} \\
	\qi(t) &= \sum_{k=0}^\infty w_k \qi_k(t), \label{qiw} \\
	r(t) &= \sum_{k=0}^\infty w_k r_k(t). \label{rw} 
\end{align}
Substituting \req{s-approx} into \req{s}, one gets
\beq{t1}
	k u^{k-1}\frac{du}{dt} = -\beta k v u^k-\alpha\beta v 
	\left(\sum_{\ell=1}^\infty \ell u^\ell
	p_\ell\right)K_0 \eta u^k+\gamma_1 \qs_k.
\eeq
Since 
\begin{align*}
\sum_{\ell=1}^\infty \ell u^\ell
p_\ell &= u \sum_{\ell=1}^\infty \ell u^{\ell-1}
p_\ell \\
&= u \frac{d\left(\sum_{\ell=1}^\infty u^\ell p_\ell\right)}{du} \\
&= u g_0'(u),
\end{align*}
Eq. \req{t1} becomes
\[
k u^{k-1}\frac{du}{dt}=-\beta k v u^k-\alpha\beta v K_0 \eta u g_0'(u)u^k+\gamma_1 \qs_k.
\]
Multiplying the preceding with $w_k$, summing from $k=0$ to infinity and
manipulating algebraically, one gets
\begin{align}
\frac{du}{dt} &= -\beta v u-\alpha\beta v K_0 \eta u g_0'(u)\frac{g_1(u)}{g_1'(u)}+
\gamma_1 \frac{\qs}{g_1'(u)}. \label{u}
\end{align}  
Similarly, we multiply the two sides of (\ref{qs})-(\ref{r}) with $w_k$ and sum
from $k=0$ to infinity.  We obtain
\begin{align}
\frac{d\qs}{dt} &= \alpha\beta K_0 \eta v u g_1(u)g_0'(u)-\gamma_1 \qs, \label{qs2} \\
\frac{dv}{dt} &= (1-\alpha)\beta v u g_1'(u)-\alpha\beta K_0 \eta v^2 u g_0'(u)
-\gamma v,\label{v2}\\
\frac{d\qi}{dt} &= \alpha\beta vug_1'(u)+\alpha\beta K_0 \eta v^2 u g_0'(u)-
\gamma \qi,\label{qi2}\\
\frac{dr}{dt} &=\gamma (v+\qi).\label{r2} 
\end{align}
Eqs. (\ref{u}), (\ref{qs2})-(\ref{r2}) form a nonlinear system of differential equations
for unknown functions $u(t)$, $\qs(t)$, $v(t)$, $\qi(t)$ and $r(t)$
subject to initial conditions $u(0), v(0), \qs(0), \qi(0)$ and $r(0)$.
In Section \ref{nsr}
we numerically solve this system and compare with the numerical solution of the system
in (\ref{s1})-(\ref{qi1}) and (\ref{r}).

Once functions $u(t)$ and $v(t)$ are obtained, we can determine
sequences of functions $s_k(t)$, $\qs_k(t)$, $x_k(t)$, $\qi_k(t)$ and 
$r_k(t)$ for $k\ge 0$ in a straight forward manner.  
First, $s_k(t)$ can be directly obtained in \req{s-approx} using $u(t)$.
Differential equations involving with $\qs_k(t)$ and $\qi_k(t)$ are linear
and can be solving using the integrating factor technique.  Function $r_k(t)$ 
can be obtained by a direct integration.

\section{Stability Analysis}\label{sa}

In this section we present a stability analysis of the nonlinear system
in (\ref{u}) and (\ref{qs2})-(\ref{r2}).  Note that once functions $u(t)$,
$\qs(t)$ and $v(t)$ are obtained, functions $\qi(t)$ and $r(t)$ can be obtained
using the technique of integrating factors and direct integration, respectively.
Thus, one just needs to consider the nonlinear system in 
(\ref{u}), (\ref{qs2}) and (\ref{v2}),
in which there are only three unknown functions.  
In order to facilitate our presentation in matrix and vector forms, we use
symbols $y_1, y_2$, and $y_3$ to denote
$u, \qs$, and $v$, respectively.  Thus, $y_1(t), y_2(t)$ and $y_3(t)$ denote
$u(t), \qs(t)$ and $v(t)$, respectively.  We rewrite (\ref{u}), (\ref{qs2}) and (\ref{v2})
using new symbols, \ie
\begin{align}
	y_1' &= -\beta y_3 y_1-\alpha\beta K_0 \eta y_3 y_1 g_0'(y_1)\frac{g_1(y_1)}{g_1'(y_1)}+
	\gamma_1 \frac{y_2}{g_1'(y_1)},\label{eq1}\\
	y_2' &= \alpha\beta K_0 \eta y_3 y_1 g_1(y_1)g_0'(y_1)-\gamma_1 y_2, \nonumber\\
	y_3' &= (1-\alpha)\beta y_3 y_1 g_1'(y_1)-\alpha\beta K_0 \eta y_3^2 y_1 g_0'(y_1)
	-\gamma y_3.\nonumber
\end{align}
Denote the right side of the equations above by $f_1, f_2$ and $f_3$, respectively.
We have
\begin{align}
	y_1' &= f_1(y_1, y_2, y_3), \nonumber\\
	y_2' &= f_2(y_1, y_2, y_3), \label{ns}\\
	y_3' &= f_3(y_1, y_2, y_3).\nonumber
\end{align} 
The preceding can be expressed in terms of vectors, \ie
\beq{ns1}
\bfx'=\bff(\bfx),
\eeq
where we use boldface letters to denote vectors and matrices.
We are interested in the stability
of equilibrium point $(y_1, y_2, y_3)=(\xi, 0, 0)$.  Recall the definition of the Jacobian
matrix of a system in the form of \req{ns1}.  The $(i, j)$ entry of the Jacobian matrix
is defined to be
\[
\left(\bfJ(\xi, 0, 0)\right)_{i,j}=\left.\frac{\partial f_i}{\partial x_j}\right|_{\bfsx=(\xi, 0, 0)}.
\]
The Jacobian matrix of the functions
on the right side of (\ref{ns}) evaluated at $(\xi, 0, 0)$ is
\begin{align}
&\bfJ(\xi, 0, 0)=\nonumber\\
&\left(\begin{array}{ccc}
	0 & \gamma_1/g_1'(\xi) & -\beta\xi-\alpha\beta K_0 \eta \xi g_0'(\xi)g_1(\xi)/
	g_1'(\xi) \\
	0 & -\gamma_1 & \alpha\beta K_0 \eta\xi g_0'(\xi)g_1(\xi) \\
	0 & 0 & (1-\alpha)\beta\xi g_1'(\xi)-\gamma\end{array}\right).\label{jm}
\end{align}
Since $\bfJ(\xi,0,0)$ is singular, the nonlinear system is not almost linear 
in the neighborhood of $(\xi, 0, 0)$.  Thus, the general stability theory of almost
linear systems can not be applied here.  In addition, in almost linear systems
equilibrium points are isolated.  In our problem, equilibrium points are not isolated.
Any point on the $y_1$ axis is an equilibrium point.  These characteristics make
each nonlinear system with singular Jacobian matrices unique.  Each problem needs a
dedicated analysis.  To see that our nonlinear system with Jacobian matrix in (\ref{jm}) 
is not almost linear, note that since
the Jacobian matrix is singular, one can apply elementary row operations to convert all
entries in the first row to zero.  Thus, the dominant terms in the first equation are not
linear.  Rather, the right side of the first equation is dominated by quadratic terms.  
 
Since probability generating functions are power series, functions $f_i, i=1, 2, 3,$ 
have continuous derivatives of all orders.  Thus, one can 
apply Taylor expansion to $f_i, i=1, 2, 3,$ around
the equilibrium point $\bfx_e=(\xi, 0, 0)^T$, where symbol $T$ denotes the
transpose operation.  For functions $f_2$ and $f_3$, we
keep only the linear terms in their Taylor expansions. For function $f_1$,
we keep both the linear term and the quadratic term.  In matrix form, we have
\beq{ns2}
\bfx'=\bfJ(\bfx_e)(\bfx-\bfx_e)+\left[\begin{array}{c}
	(\bfx-\bfx_e)^T\bfH(\bfx_e)(\bfx-\bfx_e) \\
	0 \\
	0\end{array}\right],
\eeq
where $\bfH(\bfx_e)$ is the Hessian matrix evaluated at point $\bfx_e$.  The $(i,j)$
entry of $\bfH(\bfx_e)$ is
\[
\frac{\partial^2 f_1(\bfx)}{\partial x_i \partial x_j},\qquad 1\le i, j\le 3.
\]
It is straight forward to show that the Hessian matrix of the function on the right
side of (\ref{eq1}) is 
\beq{hm}
\bfH(\bfx_e)=\left(\begin{array}{ccc}
	0 & A & B \\
	A & 0 & 0 \\
	B & 0 & 0 \end{array}\right),
\eeq
where $A$ and $B$ are defined as
\begin{align*}
	A&=\frac{-\gamma_1 g_1''(\xi)}{(g_1'(\xi))^2},\\
	B&=-\beta-\alpha\beta K_0\eta\left(g_0'(\xi)\frac{g_1(\xi)}{g_1'(\xi)}
	+\xi g_0''(\xi)\frac{g_1(\xi)}{g_1'(\xi)}+\right.\\
	&\quad\left.+\xi g_0'(\xi)\frac{(-1)g_1''(\xi)g_1(\xi)}{(g_1'(\xi))^2}+
	\xi g_0'(\xi)\right).
\end{align*}
Typically, one considers a ``translated system" of a nonlinear system in order to simplify
notations.  That is, let
\[
\tilde\bfx=\bfx-\bfx_e.
\]
Rewrite \req{ns2} in terms of $\tilde\bfx$, \ie
\beq{ts}
\tilde\bfx'=\bfJ(\bfx_e)\tilde\bfx+\left[\begin{array}{c}
	\tilde\bfx^T\bfH(\bfx_e)\tilde\bfx \\
	0 \\
	0\end{array}\right],
\eeq
Note that the equilibrium point $\bfx_e$ is translated to the origin
in the translated system.

Now we consider system \req{ts} with initial condition 
\[(\tilde y_1(0), \tilde y_2(0), \tilde y_3(0))=(\epsilon,\epsilon,\epsilon)\]
for some small positive number $\epsilon$.  Note the upper triangular form 
of the Jacobian matrix in (\ref{jm}).
We solve $\tilde x_i(t)$ successively starting from $i=3$.
From \req{ts} and (\ref{jm}), we have
\beq{eq3}
\tilde y_3'=a \tilde y_3,
\eeq
where 
\beq{a}
a=(1-\alpha)\beta \xi g_1'(\xi)-\gamma.
\eeq
Thus,
\beq{sol3}
\tilde y_3(t)=\epsilon e^{at}.
\eeq
If $a<0$, $\tilde y_3(t)\to 0$.  If $a>0$, $\tilde y_3(t)\to\infty$. 
Now we substitute \req{sol3} into the second equation in \req{ts} and obtain
\[
\tilde y_2'=-\gamma_1 \tilde y_2+\alpha\beta K_0\eta\xi g_0'(\xi) g_1(\xi) \epsilon e^{at}.
\]
This is a linear differential equation and can be solved using the technique
of integrating factors.  The solution is
\beq{sol2}
\tilde y_2(t)=(h e^{at}+(1-h) e^{-\gamma_1 t})\epsilon,
\eeq
where
\begin{align*}
	h &= \frac{\alpha\beta K_0 \eta\xi g_0'(\xi)g_1(\xi)}{a+\gamma_1}.
\end{align*}
From \req{sol2}, if $a<0$, $\tilde y_2(t)\to 0$.  If $a>0$, $\tilde y_2(t)\to\infty$.
Finally, we substitute \req{sol3} and \req{sol2} into \req{ns2}, algebraically
simplify and obtain
\beq{eq1}
\tilde y_1'=\left(d_1 e^{at}+d_2 e^{-\gamma_1 t}\right)\epsilon\tilde y_1+
\left(d_3 e^{at}+d_4 e^{-\gamma_1 t}\right)\epsilon,
\eeq
where
\begin{align*}
	d_1 &=Ah+B,\\
	d_2 &= A(1-h), \\
	d_3 &=\frac{h\gamma_1}{g_1'(\xi)}
	-\left(\beta\xi
	+\alpha\beta K_0 \eta \xi g_0'(\xi)\frac{g_1(\xi)}{g_1'(\xi)}
	\right),\\
	d_4 &=\frac{(1-h)\gamma_1}{g_1'(\xi)}.
\end{align*}
Eq. \req{eq1} is linear and its solution is
\begin{align}
	&\tilde y_1(t) = \exp\left(\frac{d_1}{a}\epsilon e^{at}-
	\frac{d_2}{\gamma_1}\epsilon e^{-\gamma_1 t}\right)
	\Biggl(\frac{\epsilon}{\exp\left(\frac{d_1\epsilon}{a}-\frac{d_2\epsilon}{\gamma_1}\right)}
	+\nonumber\\
	&\quad\int_0^t \exp\left(-\frac{d_1\epsilon}{a}e^{ay}+\frac{d_2\epsilon}{\gamma_1}
	e^{-\gamma_1 y}\right)\left(d_3\epsilon e^{ay}+d_4\epsilon e^{-\gamma_1 y}\right)\,dy\Biggr).\label{x1-sol}
\end{align}
Clearly, if $a>0$, it follows from (\ref{x1-sol}) that $\tilde y_1(t)\to\infty$.  
On the other hand, if $a<0$, as $t\to\infty$,
\begin{align}
	\tilde y_1(t) &\to\epsilon\exp\left(-\frac{d_1\epsilon}{a}+\frac{d_2\epsilon}{\gamma_1}\right)
	+\nonumber\\
	&\quad \int_0^\infty \exp\left(-\frac{d_1\epsilon}{a}e^{ay}+\frac{d_1\epsilon}{\gamma_1}
	e^{-\gamma_1 y}\right)\times\nonumber\\
	&\qquad\quad\left(d_3\epsilon e^{ay}+d_4\epsilon e^{-\gamma_1 y}\right)\,dy.\label{intg}
\end{align}
We now present an upper bound and a lower bound
for the integral on the right side of (\ref{intg}).
Since $\exp(\omega t)$ is increasing with $\omega$ for any $t$, we have
\begin{align*}
	&e^{at} \le e^{Mt}, \quad e^{-\gamma_1 t} \le e^{Mt}, \\
	&e^{at} \ge e^{mt}, \quad e^{-\gamma_1 t} \ge e^{mt},
\end{align*}
where 
\begin{align*}
	M &=\max(a, -\gamma_1),\\
	m &=\min(a,-\gamma_1).
\end{align*} 
The integral on the right side of (\ref{intg}) is bounded above
by $U\epsilon$, where
\begin{align}
	U&= \int_0^\infty \exp\left(\left|\frac{d_1\epsilon}{a}\right|e^{my}+
	\left|\frac{d_2\epsilon}{\gamma_1}
	\right|e^{m y}\right) (|d_3|+|d_4|) e^{m y}\,dy \nonumber\\
	&= \frac{|d_3|+|d_4|}{m}\frac{1-\exp\left(\left|\frac{d_1\epsilon}{a}\right|
		+\left|\frac{d_2\epsilon}{\gamma_1}\right|\right)}
	{\left(\left|\frac{d_1}{a}\right|+\left|\frac{d_2}{\gamma_1}\right|\right)\epsilon}\nonumber\\
	&\approx \frac{|d_3|+|d_4|}{|m|}.\label{U}
\end{align}
With a similar analysis, we obtain a lower bound $L\epsilon$ of the
integral on the right side of (\ref{intg}), where
\beq{L}
L=-\frac{|d_3|+|d_4|}{|m|}.
\eeq
The first term on the right side of (\ref{intg}) is approximately equal to $\epsilon$.
Thus, from (\ref{U}), it follows that 
\beq{bound}
\tilde y_1(t)\to \xi^*\in (\epsilon(1+L), \epsilon(1+U)).
\eeq
In conclusion, starting initially from $(\epsilon,\epsilon,\epsilon)$, 
the solution $(y_1(t), y_2(t), y_3(t))$ converges 
to $(\xi+\xi^*, 0, 0)$, where $\xi^*$ satisfies \req{bound}.
On the other hand, if $a > 0$, $y_i(t)\to\infty$ for $i=1, 2, 3$, as $t\to\infty$.
Note that $a$ depends only on the parameters of
the disease and the generating function of the excess degree distribution.
It does not depend on the parameters of the contact tracing and isolation policy.

\section{Numerical and Simulation Results}\label{nsr}

We present numerical and simulation results in this section.  
Numerical solution of Eqs. (\ref{r})-(\ref{qi1}) is presented in Section \ref{subsec-sde}.
We simulate the SIR model with contact tracing and isolation on five 
real-world networks.  The simulation results are shown in Section \ref{subsec-rwn}.
 
\subsection{Solution of Differential Equations}\label{subsec-sde}

\begin{figure}[htb]
	\centerline{\includegraphics[width=\linewidth]{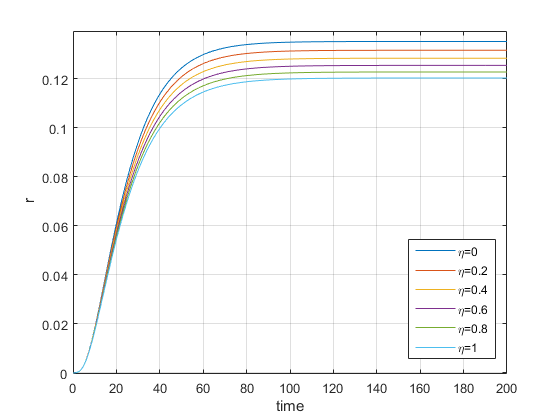}}
	\caption{Probability of recovery as a function of time for several 
	values of $\eta$.}\label{r-p-powerlaw}
\end{figure}

\begin{figure}[htb]
	\centerline{\includegraphics[width=\linewidth]{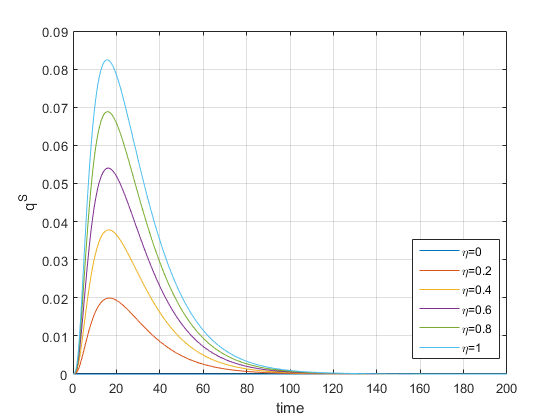}}
	\caption{Probability of isolated individuals who are susceptible 
		as a function of time for several 
		values of $\eta$.}\label{qs-p-powerlaw}
\end{figure}

\begin{figure}[htb]
	\centerline{\includegraphics[width=\linewidth]{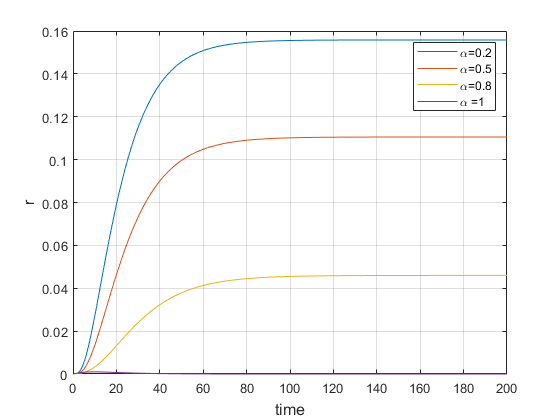}}
	\caption{Probability of recovery as a function of time for several 
		values of $\alpha$.}\label{r-a-powerlaw}
\end{figure}

\begin{figure}[htb]
	\centerline{\includegraphics[width=\linewidth]{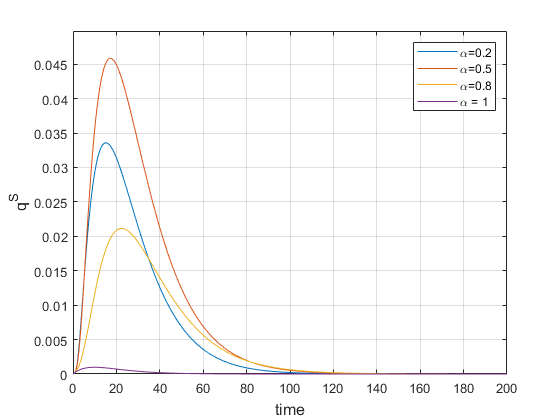}}
	\caption{Probability of isolated individuals who are susceptible 
		as a function of time for several 
		values of $\alpha$.}\label{qs-a-powerlaw}
\end{figure}

Given a degree distribution $\{p_k, k\ge 0\}$, we determine its 
corresponding excess degree distribution $\{w_k, k\ge 0\}$
according to 
\[w_k=\frac{(k+1) p_{k+1}}{K_0}\]
for $k\ge 0$ \cite{Newman2010}.  Recall that $K_0$ is the expectation of 
degree distribution $\{p_k, k\ge 0\}$.  We truncate the two degree
distributions properly such that the error is small.  
We use Matlab differential equations solver to numerically solve
Eqs. (\ref{r})-(\ref{qi1}) for Poisson degree distributions and power law
degree distributions.  We assume that the initial condition is $s_k(0)=1-10^{-3}$,
$x_k(0)=10^{-3}$ and $\qs_k(0)=\qi_k(0)=r_k(0)=0$.
We then apply \req{overall-r} to compute
the overall probability of recovery. For Poisson degree distributions,
we assume that the mean degree is 25.  For power law degree distribution, we assume
that the exponent of the distribution is $-2.5$.  We truncate both
distributions to range $[0,1000]$.  Values of other parameters are
shown in Table \ref{par-values}.  We show the probability of recovery
for the power law degree distribution in Fig. \ref{r-p-powerlaw} for several
values of $\eta$.  Note that the curve with $\eta=0$ corresponds to the 
epidemic without contact tracing and isolation.
This figure shows that the contact tracing and isolation is
effective in containing the epidemic.  In Fig. \ref{qs-p-powerlaw}
we show $\qs(t)$ for several values of $\eta$.  From this figure, we see that
while contact tracing and isolation is effective in containing the epidemic, it
comes at a cost. The contact tracing and isolation 
can be detrimental to the normal function of a society.  With a strict
isolation rule corresponding to a large value of $\eta$, we see from Fig. 
\ref{qs-p-powerlaw} that more than 8\% of the total population are
isolated.  However, these isolated individuals are susceptible.
We show how $r(t)$ and $\qs(t)$ vary with $\alpha$ in Figs. \ref{r-a-powerlaw} 
and \ref{qs-a-powerlaw}.  In the calculation of these two figures,
the value of $\eta$ is 0.5.

\begin{table}[h]
	\centering
	\begin{tabular}{|c|c|c|c|c|c|}
		\hline
		parameter & $\alpha$ & $\beta$ & $\gamma$ & $\gamma_1$ & $\eta$\\ 
		\hline
		value & 0.4 & 0.15 & 0.1 & 0.1 & 0.5 \\
		\hline
	\end{tabular}\medskip
	\caption{Parameter values used in numerical studies.}\label{par-values}
	\medskip
\end{table}

\begin{figure}[htb]
	\centerline{\includegraphics[width=\linewidth]{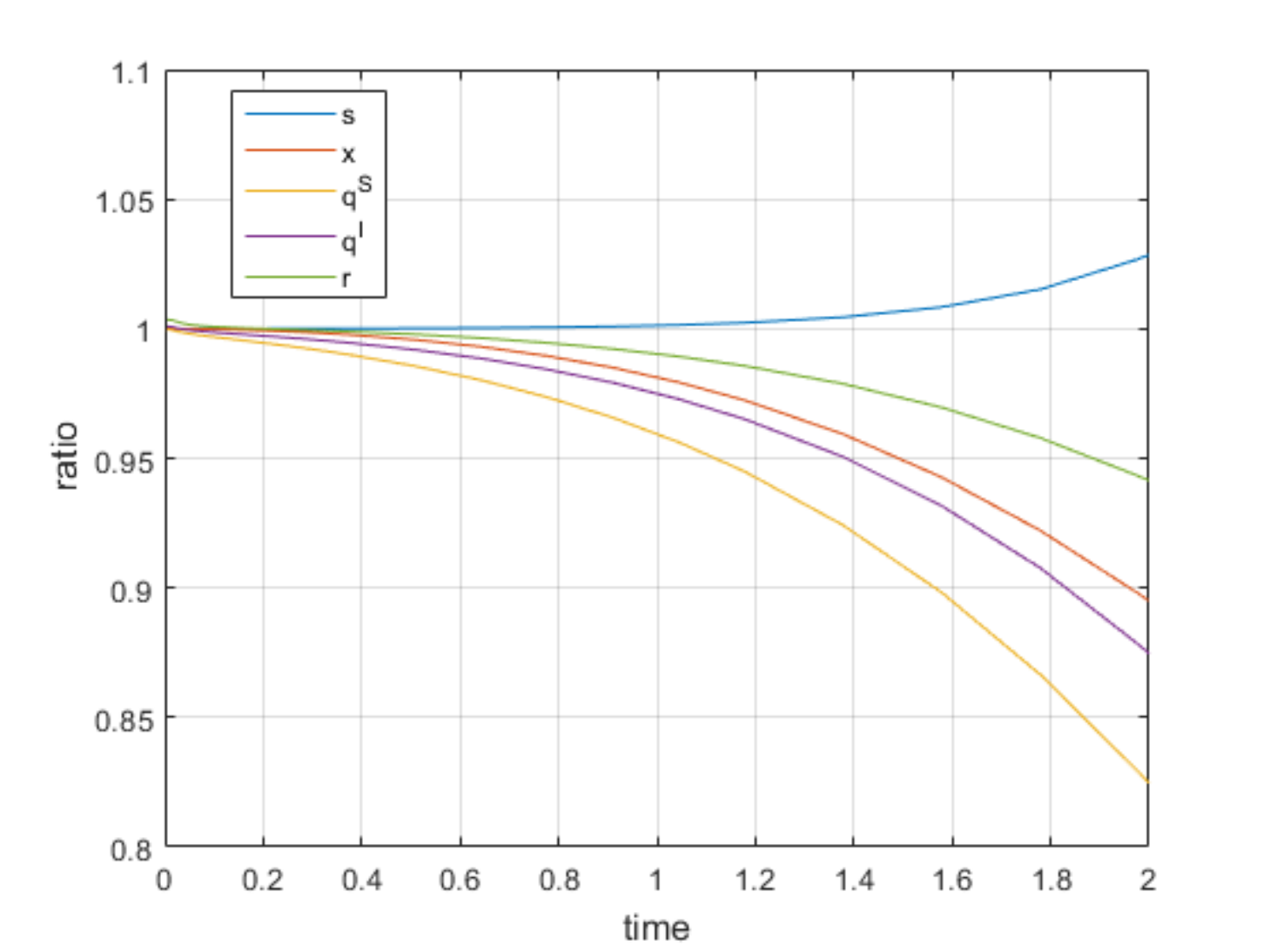}}
	\caption{Early time analysis. The ratios of early time analysis divided by
	that of the exact numerical result are shown.}\label{early-time}
\end{figure}

We now present numerical studies of the early-time analysis 
presented in Section \ref{eta}.  
We calculate a ratio by dividing the result of the early-time analysis by 
that of the exact numerical result.  The ratios are shown in Fig. \ref{early-time}.
As expected, the early time analysis is accurate when time is small and starts
to deviate when time is getting large.   

\begin{figure}[htb]
	\centerline{\includegraphics[width=\linewidth]{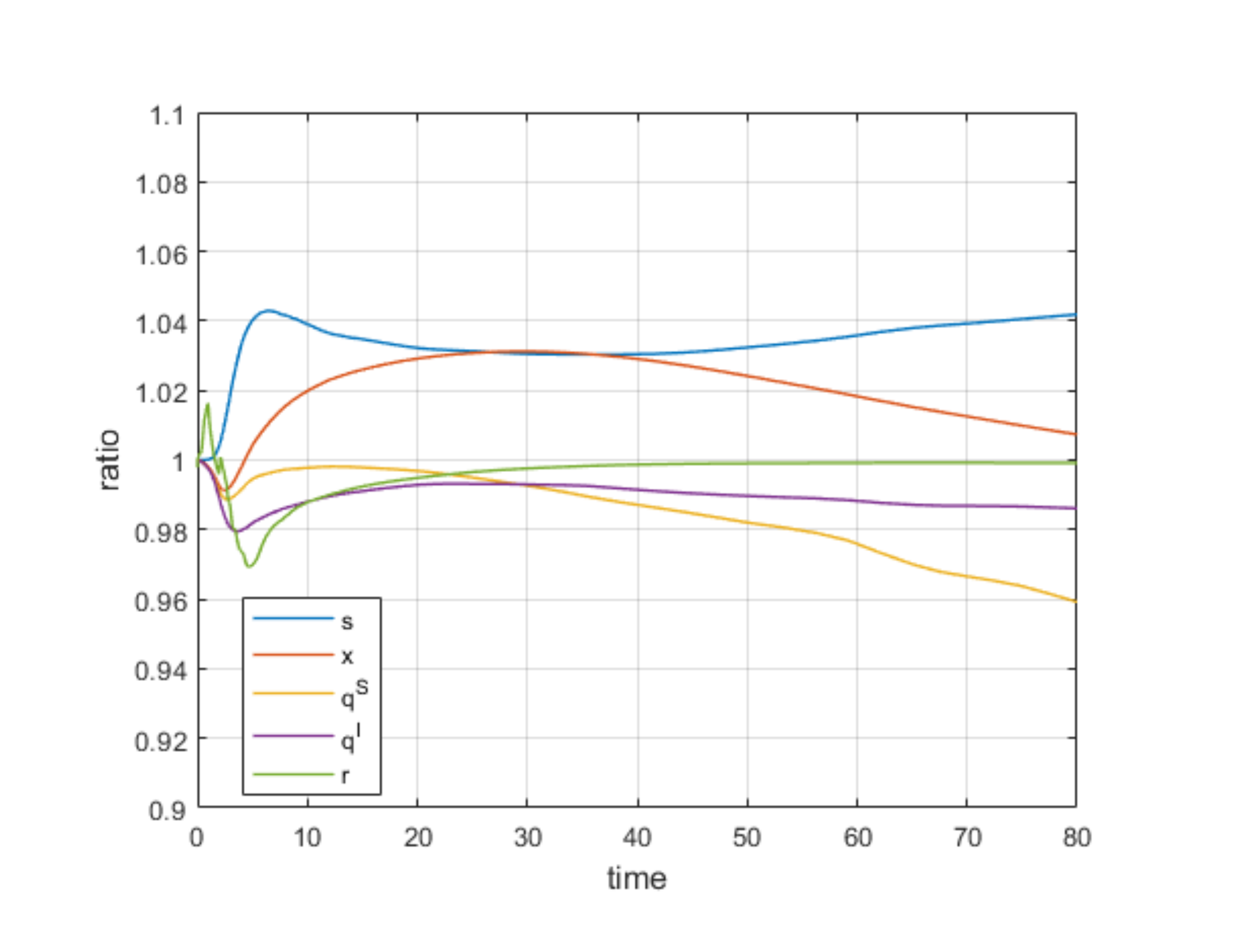}}
	\caption{Approximate analysis. The ratios of approximate analysis divided by
		that of the exact numerical result are shown.  A randomly 
		selected vertex has a Poisson degree distribution.}\label{approx-poisson}
\end{figure}

\begin{figure}[htb]
	\centerline{\includegraphics[width=\linewidth]{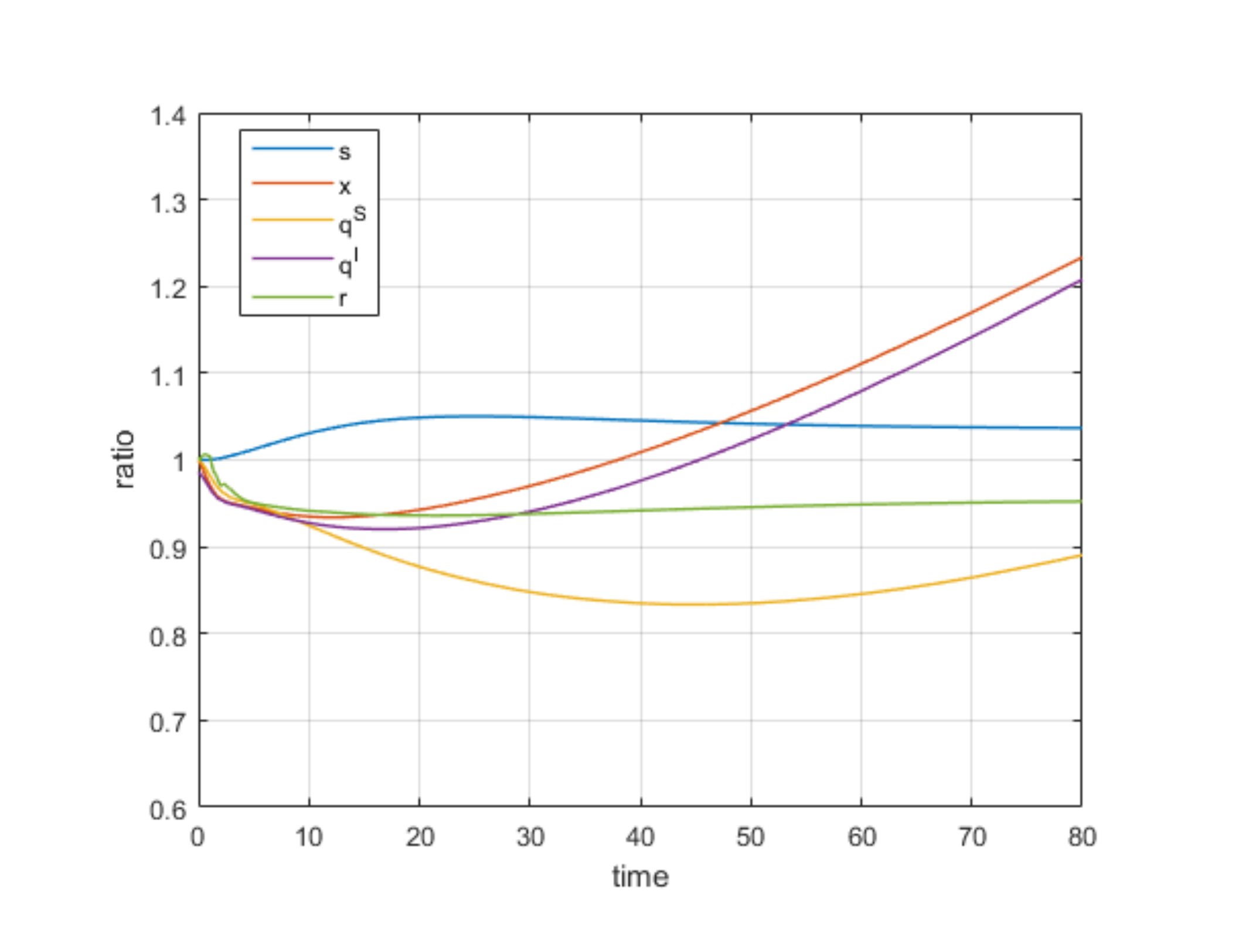}}
	\caption{Approximate analysis. The ratios of approximate analysis divided by
		that of the exact numerical result are shown.  A randomly 
		selected vertex has a power law degree distribution.}\label{approx-power-law}
\end{figure}

\begin{figure}[htb]
	\centerline{\includegraphics[width=\linewidth]{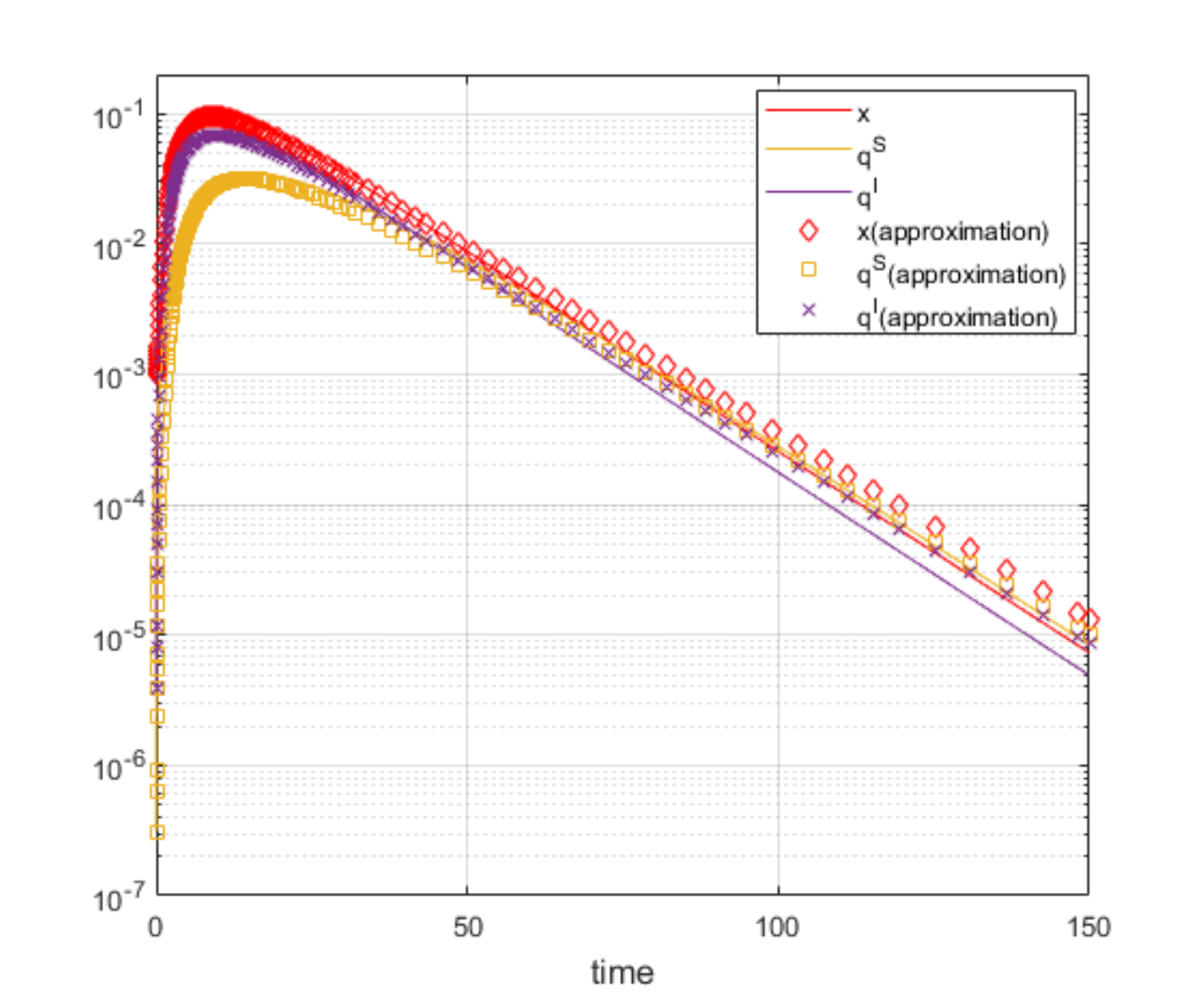}}
	\caption{Values of $x(t)$, $\qs(t)$ and $\qi(t)$ obtained by exact numerical 
		analysis and approximate analysis. Degrees have a power law  
		distribution with exponent $-2.5$.}\label{approx-qs-qi}
\end{figure}

We use Matlab to numerically solve Eqs. (\ref{u})-(\ref{r2}).  We set the initial
condition as follows.  Let $v(0)=10^{-3}$, $\qs(0)=\qi(0)=r(0)=0$, and $u(0)=g_1^{-1}(1-10^{-3})$,
where $g_1^{-1}$ is the inverse function of the probability generating function $g_1$.
We calculate a ratio by dividing the result of the approximate analysis by 
that of the exact numerical result.  The ratios are shown in Fig. 
\ref{approx-poisson}.  We see that the approximation method works quite well.
The errors are typically within five percents.  We also study the approximate
analysis of the power law degree distribution with exponent $-2.5$.    The ratios are shown in
Fig. \ref{approx-power-law}.  The errors in this case are higher, but are within
a reasonable range.  At late 
time of the epidemic, the errors corresponding to $\qs(t)$ and
$\qi(t)$ can be twenty percents. We show the exact analysis and the approximate
analysis of $x(t)$, $\qs(t)$ and $\qi(t)$ in Fig. \ref{approx-qs-qi}. 
We see that the result of exact analysis and that of the approximate analysis
are very close in early time and mid time.  At late time, the difference
between the two analyses are visible.  However, the values of 
$x(t)$, $\qs(t)$ and $\qi(t)$ are very small and they are less significant
to the epidemics.

\subsection{Simulation of Real-world Networks}\label{subsec-rwn}

\begin{figure}[htb]
	\centerline{\includegraphics[width=0.9\linewidth]{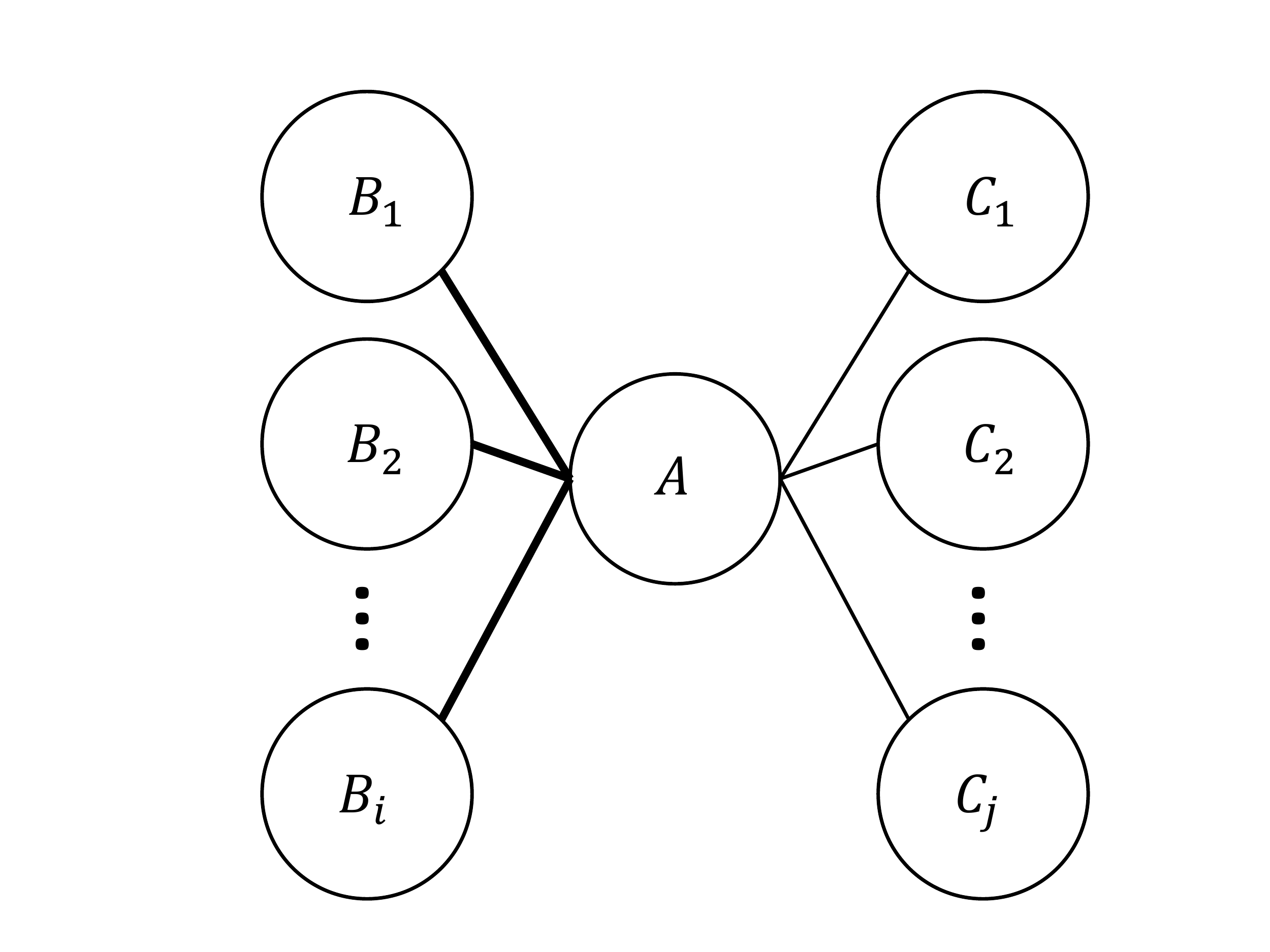}}
	\caption{Vertex $A$ has $i$ close contacts denoted by $B_1, B_2, \ldots, B_i$
		and $j$ normal contacts denoted by $C_1, C_2, \ldots, C_j$.
		The edges between vertex $A$ and $B_1, B_2, \ldots, B_i$ are
		the first type of edges. The edges between vertex $A$ and 
		$C_1, C_2, \ldots, C_j$ are the second type of edges. The edges connecting
		close contacts are denoted by thick lines.}\label{close-contacts}
\end{figure}

We analyze contact tracing and isolation using configuration network models.
Configuration models are mathematically simple, but lack some important
characteristics such as clustering and degree correlations.  It is known
that real-world networks possess these characteristics \cite{Newman2010}
and it is known that these characteristics have non-neglectable impacts to
many networking problems \cite{Zhang2022}.
It is well known that a group of 
individuals with close contacts with each other, such as members in a 
household or workmates in an office, are likely to contract the disease
if one in the group does \cite{House.2008,House.2009}.  
In this section we study the SIR model with contact tracing and isolation
on real-world networks by simulation.  We simulate a more general contact
tracing and isolation policy.  In our simulation
model there are two types of edges.  The first type of edges connects
two close contacts.  The disease has a higher transmission rate 
across edges across the first type of edges. The second type of edges
connects two normal contacts.  We refer the reader to Fig. 
\ref{close-contacts} for a graphical illustration.   Suppose that vertex
$A$ is infected and shows symptoms.   We assume that the contact tracing
and isolation policy quarantines and isolates all $A$'s close contacts, \ie
$B_1, B_2, \ldots, B_i$.  In addition, the policy isolates a fraction $\eta$
of $A$'s normal contacts.  Specifically, each $C_k$, for $1\le k\le j$, 
is isolated with probability $\eta$.  

To the best of our efforts, we have not been able to find
datasets of real-world networks, in which edges are marked according to
whether they connect two close contacts or two normal contacts.  
Tie strength is a fundamental topic in sociology \cite{Granovetter1975}.
Strong ties generally refer to edges corresponding to closer friendships or greater
frequency of interactions.  On the other hand, weak ties generally
refer to edges connecting two acquaintances \cite{Easley2010}. 
Given a social graph, there have been efforts to mark the strengths
of the edges in the graph \cite{Onnela2007, Marsden1984}.
Particularly, Onnela et al. \cite{Onnela2007} studied a network formed
by communications among cellular phones over an 18-week period.
Onnela et al. found tie strengths and {\em neighborhood overlap} are 
highly correlated. The neighborhood overlap of an edge connecting 
vertices $A$ and $B$ is defined as the ratio
\beq{def-no}
\frac{|N_A \cap N_B|}{|N_A\cup N_B - \{A, B\}|},
\eeq
where $N_A$ and $N_B$ denote the sets of $A$ and $B$'s neighbors, respectively.
The denominator of \req{def-no} is the number of $A$'s neighbors and 
$B$'s neighbors.  However, $A$ and $B$ are excluded.
Clearly, the neighbor overlap defined in \req{def-no} is a number 
between zero and one.  In our experiments, we mark an edge as a type one
edge if its neighbor overlap is more than or equal to a threshold $h$.
Otherwise, the edge is marked as a type two edge.

We consider five networks collected in the real world.  Their information
is summarized in Table \ref{rwn-info}. The bitcoin and the facebook datasets
are available at the SNAP network datasets site \cite{snapnets}.  
The dolphin, tvshow and the anybeat datasets are available at 
Network Data Repository \cite{nr}. We simulate the SIR model with contact
tracing and isolation in these networks.  The results are presented in
Table \ref{seir-rwn-1} and Table \ref{seir-rwn-2}.
Quarantine period is the number of time units that one remains isolated.  $S$ 
(resp. $R$) denotes the fraction of population that are susceptible (resp. recovered) at the 
end of the epidemic.  $Q_{max}$ denotes the maximum fraction 
of population that are isolated and are susceptible (resp. infected) during 
the entire process of epidemic.  $I_{max}$ is the maximum fraction of
asymptomatic individuals who are not isolated during the epidemic.
$t_q$ (resp. $t_i$) is the amount of time
for the number of quarantined individuals (resp. number of infected individuals) to 
reduce to zero.  Among all neighbors who are connected
by weak ties with an infected individuals, a fraction of $\eta$ of neighbors are 
isolated.  In Table \ref{seir-rwn-1} two values of
$\eta$ were simulated, and in Table \ref{seir-rwn-2} two values of $\eta$ were simulated.
We draw the following conclusions from these two tables.
\begin{enumerate} 
\item If one tightens the contact tracing and isolation rules by extending 
the isolation period or increasing $\eta$, less people are 
infected with the disease. Simulation shows that $S$ (resp. $R$) is increasing 
(resp. decreasing) with the isolation period.
\item With a more stringent isolation policy enforced, there are more susceptible
individuals who are isolated, and there are less asymptomatic individuals who are
not isolated. Simulation shows that $Q_{max}$ (resp. $I_{max}$)
is increasing (resp. decreasing) with isolation period and $\eta$.
\item With a more stringent isolation policy enforced, it takes longer to contain 
the epidemic.  Simulation shows that both $t_q$ and $t_i$ are increasing with 
the isolation period.
\item With the same isolation period and the same value of $\eta$, the contract tracing
and isolation policy works more efficiently if a larger fraction $h$ of neighbors 
are close contacts.   
\end{enumerate}

\begin{table}[htb]\bigskip
	\centering
		\begin{tabular}{|c|c|c|c|c|c|}
			\hline
			data set& $n$ & $m$ & $K_0$ & $\rho$ & $C$\\ 
			\hline\hline
			dolphin & 62 & 159 & 5.13 & -0.0436 & 0.2590\\
			\hline
			bitcoin & 5881 & 21492 & 7.31 & -0.1648 & 0.1775\\
			\hline
			tvshow & 3892 & 17262 & 8.87 & 0.5604& 0.3737\\
			\hline
			facebook & 4039 & 88234 & 43.69 &0.0636 & 0.6055\\
			\hline
			anybeat & 12645 & 49132 & 7.77 &-0.1234 & 0.2037\\
			\hline
		\end{tabular}
	\medskip
\caption{Five real-world networks. $n$, $m$, $K_0$ are the number of
		vertices, the number of edges and the expected degree of a randomly selected
		vertex of the network.  $\rho$ is the Pearson degree correlation of a randomly selected
		edge, and $C$ is the clustering coefficient.}\label{rwn-info}
	\medskip
\end{table}

\newlength\q
\setlength\q{\dimexpr .5\textwidth -2\tabcolsep}
\noindent

\begin{table*}[h!]
	\centering
	\begin{threeparttable}
		\begin{tabular}{|c|l||c|c|c|c|c|c||c|c|c|c|c|c|} 
			\hline
			\multirow{2}{*}{data set}&isolation& \multicolumn{6}{c|}{$\eta=0.6$}& \multicolumn{6}{c|}{$\eta=0.9$}\\
			\cline{3-14}
			&period&$S$ & $R$ & $Q_{max}$ & $I_{max}$ & $t_{q}$ & $t_{i}$& $S$ & $R$ & $Q_{max}$ & $I_{max}$ & $t_{q}$ & $t_{i}$\\
			\hline\hline
			\multirow{3}{*}{dophin}&3	&0.734 	&0.266 	&0.087 	&0.090 	&28 	&38 &0.752 	&0.248 	&0.111 	&0.091 	&27 	&38 \\ \cline{2-14}
			&7 							&0.737 	&0.263 	&0.104 	&0.090 	&30 	&38 &0.777 	&0.223 	&0.120 	&0.083 	&27 	&36 \\ \cline{2-14}
			&14							&0.755 	&0.246 	&0.114 	&0.085 	&32 	&38 &0.790 	&0.210 	&0.137 	&0.081 	&30 	&35 \\ 
			\hline						
			\multirow{3}{*}{bitcoin}&3	&0.635 	&0.365 	&0.161 	&0.095 	&93 	&101 &0.657 	&0.343 	&0.210 	&0.090 	&95 	&105 \\ \cline{2-14}
			&7 							&0.631 	&0.369 	&0.203 	&0.093 	&101 	&109 &0.657 	&0.343 	&0.257 	&0.086 	&104 	&113 \\ \cline{2-14}
			&14							&0.640 	&0.360 	&0.214 	&0.090 	&115 	&123 &0.663 	&0.337 	&0.263 	&0.085 	&120 	&129 \\
			\hline						
			\multirow{3}{*}{tvshow}&3	&0.603 	&0.397 	&0.055 	&0.080 	&108 	&119 &0.658 	&0.342 	&0.066 	&0.071 	&105 	&118 \\ \cline{2-14}
			&7 							&0.622 	&0.378 	&0.085 	&0.072 	&113 	&124 &0.681 	&0.319 	&0.099 	&0.062 	&111 	&123 \\ \cline{2-14}
			&14							&0.658 	&0.342 	&0.115 	&0.065 	&120 	&128 &0.720 	&0.280 	&0.129 	&0.057 	&117 	&127 \\ 
			\hline						
			\multirow{3}{*}{facebook}&3	&0.202 	&0.798 	&0.144 	&0.319 	&89 	&103 &0.238 	&0.762 	&0.179 	&0.300 	&91 	&107 \\ \cline{2-14}
			&7 							&0.212 	&0.788 	&0.196 	&0.309 	&96 	&109 &0.255 	&0.745 	&0.238 	&0.289 	&99 	&114 \\ \cline{2-14}
			&14							&0.238 	&0.762 	&0.227 	&0.307 	&110 	&120 &0.284 	&0.716 	&0.268 	&0.287 	&114 	&129 \\
			\hline						
			\multirow{3}{*}{anybeat}&3	&0.673 	&0.327 	&0.190 	&0.115 	&91 	&100 &0.690 	&0.310 	&0.245 	&0.114 	&90 	&100 \\ \cline{2-14}
			&7 							&0.673 	&0.327 	&0.209 	&0.117 	&95 	&104 &0.697 	&0.303 	&0.274 	&0.112 	&95 	&106 \\ \cline{2-14}
			&14							&0.679 	&0.321 	&0.213 	&0.115 	&102 	&111 &0.698 	&0.302 	&0.282 	&0.114 	&105 	&114 \\
			\hline
		\end{tabular}
	\end{threeparttable}\medskip
\caption{Simulation results of the SIR model with contact tracing 
	on five real-world networks. Two values of $\eta$ were simulated.  
	The value of $h$ in this set of 
	simulation is $0.75$.}\label{seir-rwn-1}\medskip
\end{table*}

\begin{table*}[h!]
	\centering
	\begin{threeparttable}
		\begin{tabular}{|c|l||c|c|c|c|c|c||c|c|c|c|c|c|} 
			\hline
			\multirow{2}{*}{data set}&isolation& \multicolumn{6}{c|}{$h=0.3$}& \multicolumn{6}{c|}{$h=0.6$}\\
			\cline{3-14}
			&period&$S$ & $R$ & $Q_{max}$ & $I_{max}$ & $t_{q}$ & $t_{i}$& $S$ & $R$ & $Q_{max}$ & $I_{max}$ & $t_{q}$ & $t_{i}$\\
			\hline\hline
			\multirow{3}{*}{dolphin}	&3	&0.667 	&0.333 	&0.082 	&0.112 	&32 	&42 	&0.692 	&0.308 	&0.060 	&0.092 	&32 	&41 \\ \cline{2-14}
			&7	&0.674 	&0.326 	&0.097 	&0.109 	&32 	&41 	&0.709 	&0.291 	&0.072 	&0.086 	&33 	&40 \\ \cline{2-14}
			&14	&0.706 	&0.294 	&0.108 	&0.101 	&34 	&40 	&0.724 	&0.276 	&0.078 	&0.084 	&34 	&39 \\ 
			\hline
			\multirow{3}{*}{bitcoin}	&3	&0.628 	&0.372 	&0.124 	&0.092 	&96 	&105 	&0.600 	&0.400 	&0.107 	&0.103 	&94 	&102 \\ \cline{2-14}
			&7	&0.627 	&0.373 	&0.156 	&0.088 	&103 	&112 	&0.602 	&0.398 	&0.138 	&0.096 	&98 	&106 \\ \cline{2-14}
			&14	&0.634 	&0.366 	&0.164 	&0.085 	&118 	&125 	&0.606 	&0.394 	&0.147 	&0.097 	&107 	&115 \\
			\hline												
			\multirow{3}{*}{tvshow}	&3	&0.546 	&0.454 	&0.041 	&0.096 	&105 	&115 	&0.530 	&0.470 	&0.038 	&0.090 	&109 	&118 \\ \cline{2-14}
			&7	&0.562 	&0.438 	&0.064 	&0.088 	&108 	&116 	&0.543 	&0.457 	&0.060 	&0.083 	&112 	&120 \\ \cline{2-14}
			&14	&0.590 	&0.410 	&0.090 	&0.084 	&114 	&123 	&0.563 	&0.437 	&0.087 	&0.078 	&119 	&126 \\
			\hline										
			\multirow{3}{*}{facebook}	&3		&0.156 	&0.844 	&0.091 	&0.352 	&86 	&99 	&0.152 	&0.848 	&0.096 	&0.338 	&87 	&99 \\ \cline{2-14}
			&7	&0.168 	&0.832 	&0.131 	&0.348 	&89 	&101 	&0.163 	&0.837 	&0.134 	&0.330 	&92 	&102 \\ \cline{2-14}
			&14	&0.189 	&0.811 	&0.152 	&0.347 	&97 	&107 	&0.182 	&0.818 	&0.155 	&0.328 	&99 	&109 \\
			\hline										
			\multirow{3}{*}{anybeat}	&3	&0.641 	&0.359 	&0.116 	&0.131 	&89 	&97 	&0.641 	&0.359 	&0.113 	&0.121 	&91 	&99 \\ \cline{2-14}
			&7	&0.644 	&0.356 	&0.132 	&0.132 	&90 	&99 	&0.644 	&0.356 	&0.128 	&0.120 	&93 	&101 \\ \cline{2-14}
			&14	&0.650 	&0.350 	&0.137 	&0.131 	&95 	&104 	&0.651 	&0.349 	&0.131 	&0.119 	&97 	&105 \\		
			\hline
		\end{tabular}	
	\end{threeparttable}\medskip
\caption{Simulation results of the SIR model with contact tracing 
	on five real-world networks. Two values of $h$ were simulated.  
	The value of $\eta$ in this set of 
	simulation is $0.6$.}\label{seir-rwn-2}\bigskip
\end{table*}

\section{Conclusions}\label{conclusions}
In this paper we have presented a degree based approximation to the
SIR model with contact tracing and isolation.  We proposed an approximation
method which greatly reduced the numerical complexity in solving the
differential equations in the degree based approximation.  
We have also simulated the SIR model with contact tracing and isolation
on five real-world networks.  We showed that
contact tracing and isolation are effective to control epidemics.

\bibliography{bibdatabase}
\bibliographystyle{IEEEtran}

\end{document}